\documentclass[11pt,twoside]{article}
\usepackage{asp2010}

\resetcounters

\begin{document}

\title{Chromospheric magnetic fields. \\ Observations, simulations and their interpretation}
\author{J. de la Cruz Rodr\'iguez$^1$, H. Socas-Navarro$^2$, M. Carlsson$^3$, J. Leenaarts$^3$
\affil{$^1$ Department of Physics and Astronomy, Uppsala University, Box 516, SE-75120 Uppsala, Sweden}
\affil{$^2$   Instituto de Astrof\'isica de Canarias, Avda V\'ia L\'actea S/N, 
  La Laguna 38205, Tenerife, Spain}
\affil{$^3$ Institute of Theoretical Astrophysics, University of Oslo, 
  P.O. Box 1029 Blindern, N-0315 Oslo, Norway}
}

\begin{abstract}
 The magnetic field of the quiet-Sun chromosphere remains a mystery for solar physicists. The reduced number of chromospheric lines are intrinsically hard to model and only a few of them are magnetically sensitive. 
In this work, we use a 3D numerical simulation of the outer layers of the solar atmosphere, to asses the reliability of non-LTE inversions, in this case applied to the \ion{Ca}{ii}~$\lambda8542$~\AA \ line. We show that NLTE inversions provide realistic estimates of physical quantities from synthetic observations.

\end{abstract}

\section{Introduction}\label{sec:intro}

During the past decades, the \ion{Ca}{ii}  lines have been used to study the solar atmosphere. The strong H \& K lines and the infrared triplet have been particularly popular in chromospheric studies \citep[see][and references therein]{1974SoPh...39...49S,1991SoPh..134...15R,2000Sci...288.1396S,2001ApJ...552..871L,2006ApJ...640.1142P,2009ApJ...694L.128L,2011arXiv1110.6606R}. 

The interest for the infrared lines has grown during the past decade as un-precedented observations were carried-out \citep{2008A&A...480..515C} and new diagnosing tools were developed \citep{2000ApJ...530..977S}. During the past 10 years, the development of a new generation of Fabry-P\'erot interferometers has contributed to the acquirement of high-resolution observations over a relatively large field-of-view on the Sun, with a restricted wavelength coverage, using the \ion{Ca}{ii}~$\lambda8542$~\AA \ line, making chromospheric studies even more attractive.

\begin{figure}[]
      \centering
        \resizebox{\textwidth}{!}{\includegraphics[]{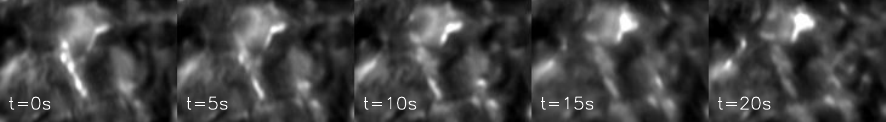}}
        \caption{Time evolution of a $10.3\arcsec \times 7.2\arcsec$ patch of quiet Sun at the core of the \ion{Ca}{ii}~$\lambda8542$~\AA \ line.}
        \label{fig:quiet_tseries}
\end{figure}

To achieve high polarimetric sensitivity, long exposure times are needed. However, chromospheric dynamics limit the the exposure time in chromospheric observations, especially if image reconstruction is needed \citep{2006ApJ...648L..67V}. To illustrate this problem, a time series acquired with 5 seconds cadence is illustrated in Fig.~\ref{fig:quiet_tseries}. The difference between consecutive images is large enough to show chromospheric motions in very small time scales.

Additionally, the relatively low effective Land\'e factor $\mathrm{g}_\mathrm{eff} = 1.10$ of the \ion{Ca}{ii}~$\lambda8542$~\AA \ line, and the wide Gaussian core (a common feature in all the lines of the \ion{Ca}{ii}~IR triplet) gives low polarization levels compared to the narrower \ion{Fe}{i} lines, commonly used in photospheric spectropolarimetry ($\mathrm{g}_\mathrm{eff} >2$).

Unfortunately, scattering polarization also must be considered to interpret quiet-sun observations \citep{2010ApJ...722.1416M,2010mcia.conf..118T}.  For these reasons, most attempts to measure, reconstruct or understand the magnetic topology of the chromosphere have focused on active regions and solar network \citep{2000Sci...288.1396S,2001ApJ...552..871L,2007ApJ...663.1386P} and off-limb spicules \citep{2005ApJ...619L.191T,2005A&A...436..325L,2010ApJ...708.1579C}. 

The \ion{Ca}{ii}~$\lambda8542$~\AA \ line cannot be modeled assuming local thermodynamic equilibrium (LTE hereafter). However, the statistical equilibrium equations (one of the simplest non-LTE recipes) can be used to compute the model-atom population densities \citep{2011A&A...528A...1W}. Furthermore, in active regions, it is reasonably safe to assume Zeeman induced polarization. \citet{2000ApJ...530..977S} presented a scheme to carry-out NLTE inversions of spectropolarimetric data, assuming statistical equilibrium and Zeeman polarization.

Two datasets acquired with the Swedish 1-m Solar Telescope \citep[SST,][]{2003SPIE.4853..341S} and the CRIsp SpectroPolarimeter \citep[CRISP,][]{2006A&A...447.1111S}, are described below illustrating the complexity of the observational data.

\begin{figure}[]
      \centering
        \resizebox{\textwidth}{!}{\includegraphics[]{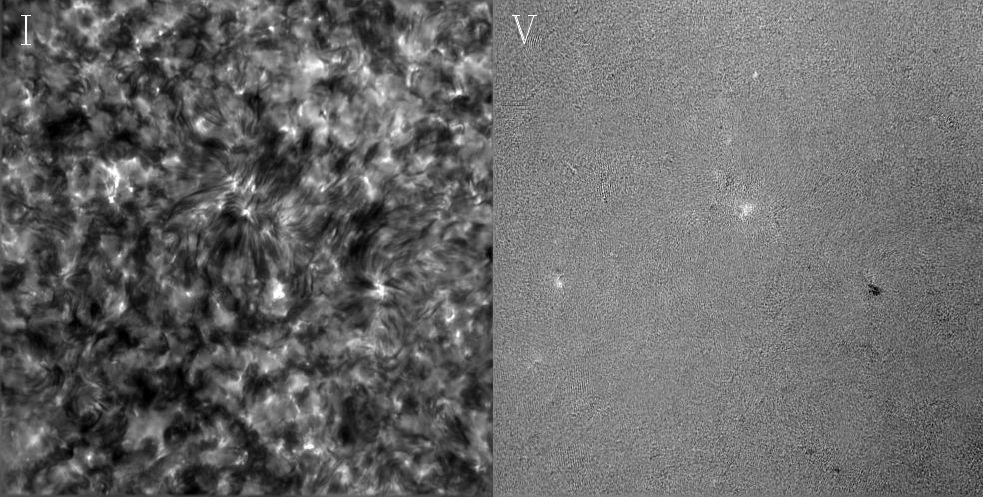}}
        \caption{Very quiet Sun observation at -231 m\AA \ from the core of the \ion{Ca}{ii}~$\lambda8542$ line. Stokes $I$ and $V$ are shown on the left and right panels respectively. The data were acquired on 2009-05-29 and the field size is $70\arcsec \times 70\arcsec$.}
        \label{fig:quiet_pol}
\end{figure}

Figure~\ref{fig:quiet_pol} shows very quiet-Sun data at $-231$ m\AA \ from the core of \ion{Ca}{ii}~$\lambda8542$. The \textit{clapotisphere} described in \citet{2011arXiv1110.6606R} is clearly visible all over the field-of-view. Fibril-like features originate from bright points located in the center of the image. The weak signal in the Stokes~$V$ image originates in bright points and large-scale patterns are not visible. The signal in Stokes~$Q$ and $U$ is, as expected, below the noise level.

\begin{figure}[]
      \centering
        \resizebox{\textwidth}{!}{\includegraphics[]{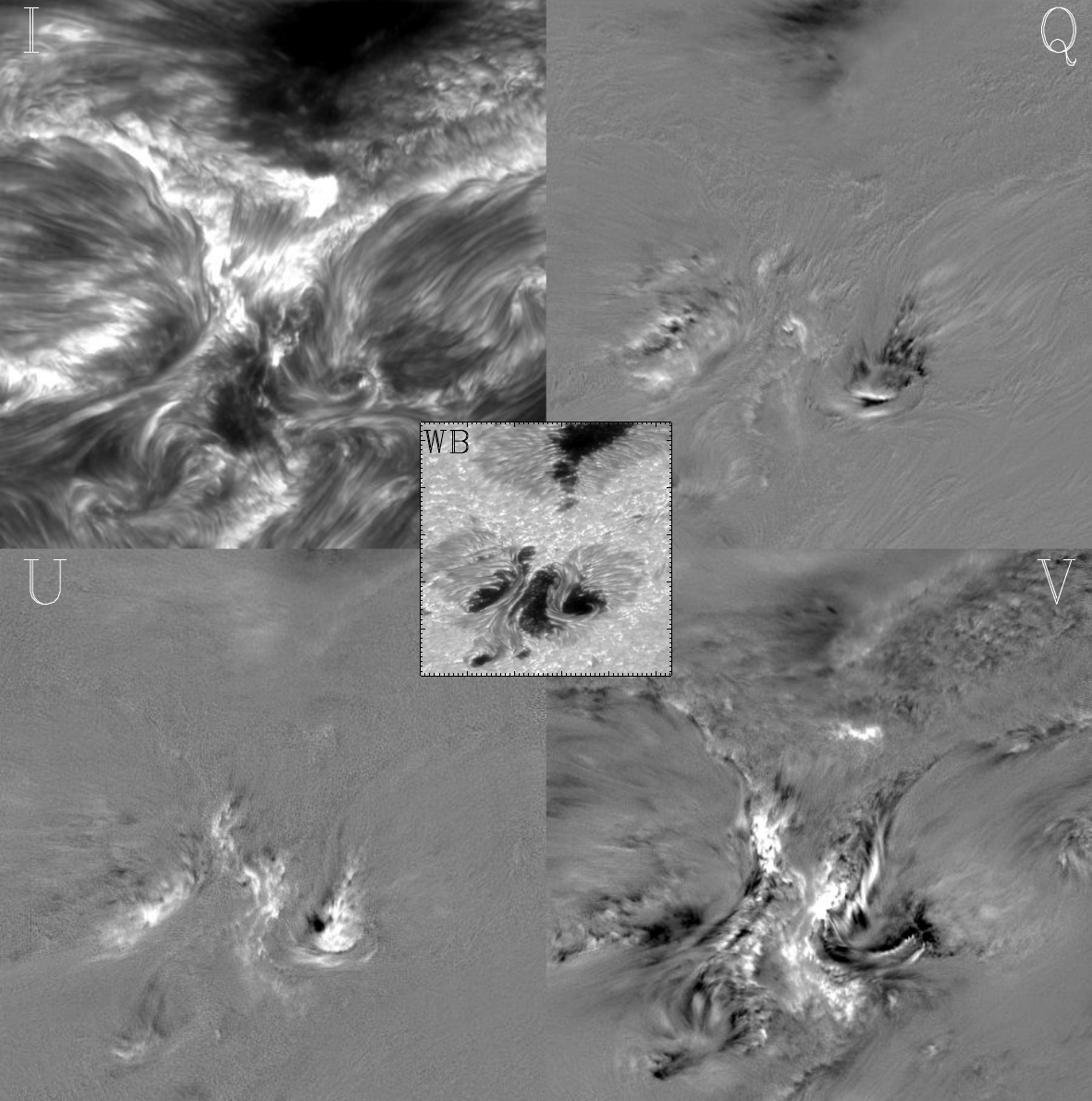}}
        \caption{Active region observed at -225 m\AA \ from the core of the \ion{Ca}{ii}~$\lambda8542$ line. Stokes $I$, $Q$, $U$, $V$ are shown from left to right and top to bottom respectively. A photospheric image is shown in the central panel to set the context, where the tick mark separation is $10\arcsec$ and the total FOV $53\arcsec \times 53\arcsec$. Data processing by J. de la Cruz Rodr\'iguez and L. Rouppe van der Voort. Observations by A. Ortiz-Carbonell and B. Hole (2011-09-24). }
        \label{fig:active_pol}
\end{figure}

Active regions show a completely different scenario. The Stokes~$Q$, $U$ and $V$ panels in Fig.~\ref{fig:active_pol} show areas with strong signal. The complicated features seen in Stokes~$V$ are not all due to the magnetic field topology, but to Doppler motions and emission in the Stokes~$I$ profile, which can lead to a reversal of the $V$ profile.

The work presented here attempts to asses the reliability of NLTE inversions and how to interpret the inversion results. For that purpose we have created synthetic observations from a 3D simulation of the solar chromosphere, evaluating the radiation field in 3D. Those synthetic profiles are degraded with Gaussian instrumental profiles and random noise before inverting them. The resulting magnitudes are compared with those from the original 3D snapshot.

NLTE inversions have been used in the past to infer physical quantities in the solar chromosphere \citep[see][]{2000ApJ...530..977S, 2007ApJ...670..885P}, however, the work partially presented in this paper aims to asses the reliability of the results and evaluate possible systematic errors induced by the physical assumptions used during the inversion process, some of them listed below:

\begin{itemize}
	\item 3D NLTE effects: the radiation field propagates through the 3D structure of the solar atmosphere, however, the inversion decouples it, treating each column as a plane-parallel atmosphere (usually referred to as 1.5D). 
	\item Hydrostatical equilibrium and the ideal gas equation of state are assumed to allow physical consistency between electron pressure, electron density and temperature.
\end{itemize} 

To evaluate these two effects, we create realistic synthetic observations from a snapshot from a 3D numerical simulation from the photosphere to the corona. The inversions are carried out using our synthetic observations, previously degraded assuming a list of instrumental profiles and noise.

\section{Radiative transfer}\label{sec:rt}
This work is based on that carried out by \citet{2009ApJ...694L.128L}. To create \textit{realistic} observations we use a 3D simulation computed with the Oslo Stagger Code \citep[see][]{2007ASPC..368..107H}. The snapshot extends from the upper convection zone to the corona, and covers a physical patch on the surface of the Sun of $16.6\times8.3\times15.5$ Mm ($256\times128\times160$~pixels).

The snapshot has been interpolated to the higher depth-resolution but restricted to the zone of importance for the radiative transfer calculations. The simulation has an average magnetic field strength of 150 G in the photosphere.

The NLTE radiative transfer calculations are performed with a \ion{Ca}{ii} model atom consisting of 5 bound levels plus a continuum. The synthetic observations are computed in two steps. To compute the populations densities we use the radiative transfer code \textsc{Multi3D} \citep{2009ASPC..415...87L} that allows to evaluate the 3D radiation field (3D~NLTE hereafter). The resulting population densities are used by \textsc{Nicole} (see~H.~Socas-Navarro contribution) to compute the full-Stokes vector. The full-Stokes profiles are normalized to the average continuum intensity.

\begin{figure}[]
      \centering
        \resizebox{\textwidth}{!}{\includegraphics[trim=0cm 0.7cm 0.1cm 0cm, clip=true]{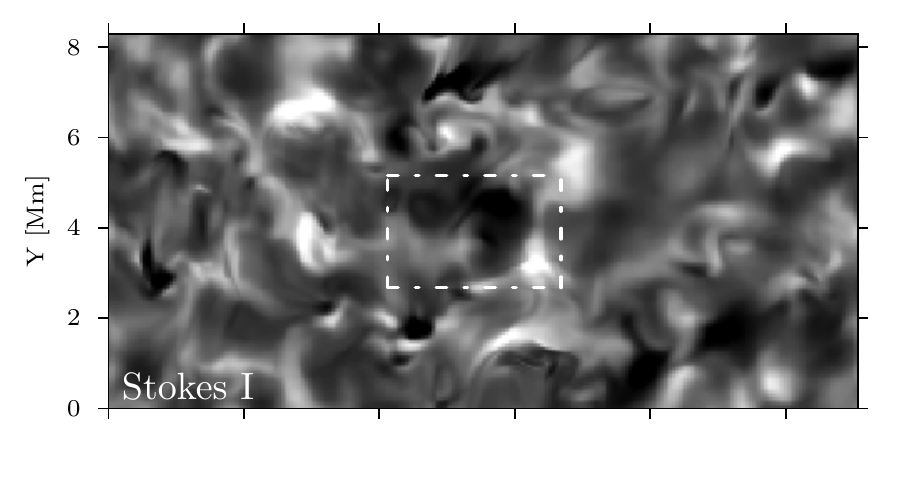}\includegraphics[trim=1cm 0.7cm 0.2cm 0cm, clip=true]{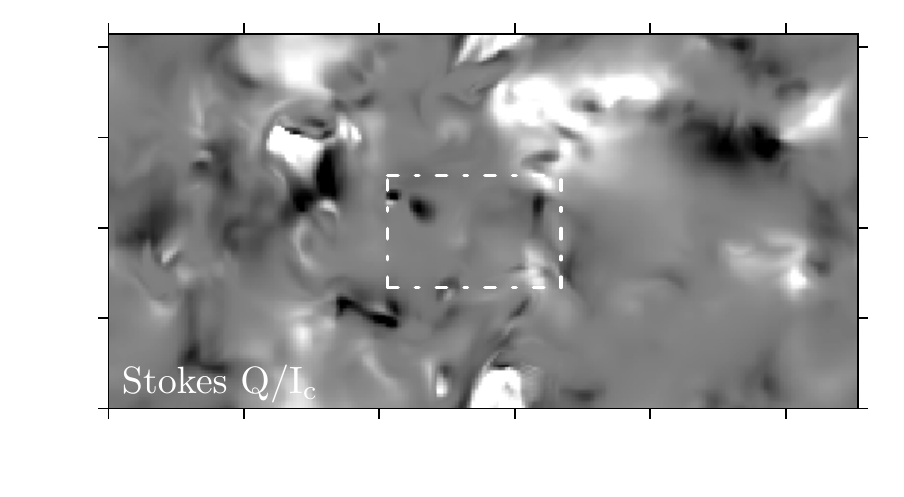}}
        \resizebox{\textwidth}{!}{\includegraphics[trim=0.cm 0cm 0.1cm 0.2cm, clip=true]{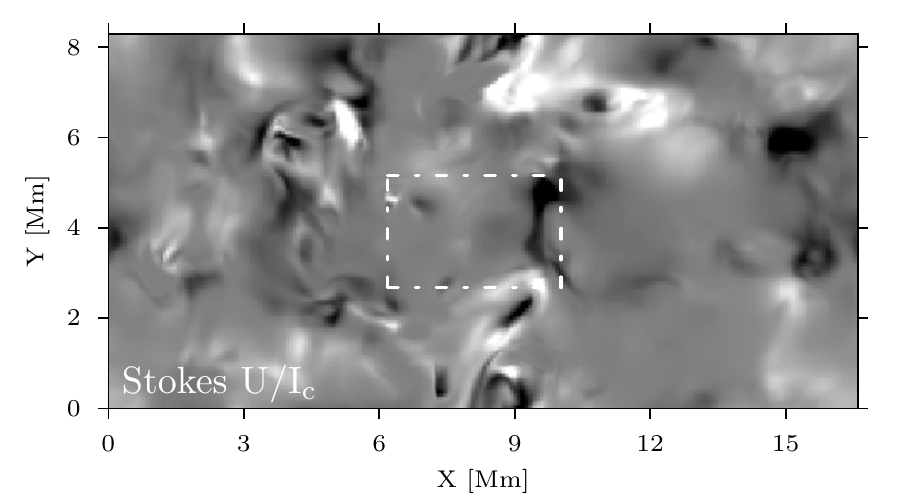}\includegraphics[trim=1cm 0cm 0.2cm 0.2cm, clip=true]{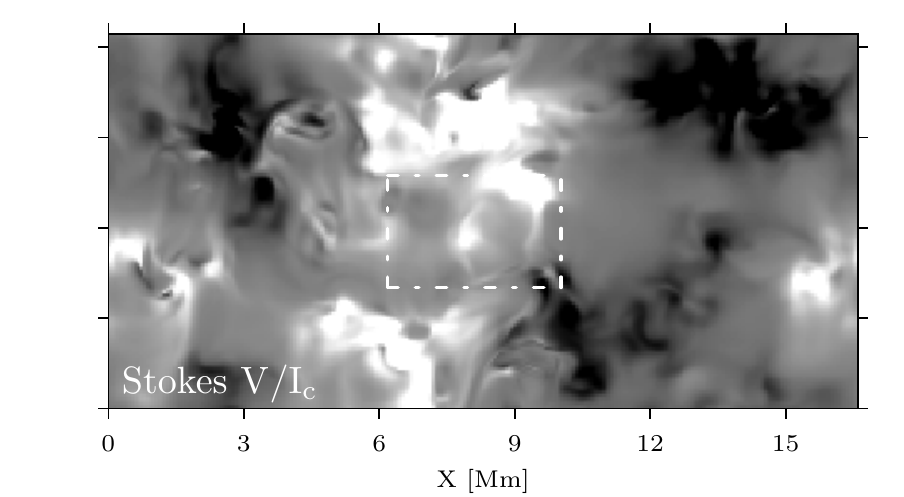}}

        \caption{Synthetic observations at 209 m\AA \ from the core of the \ion{Ca}{ii}~$\lambda8542$ \AA \ line. The Stokes $Q$ and $U$ images are scaled between $\pm 6 \cdot 10^{-4}$ and Stokes $V$ is scaled between $\pm 1.5\cdot 10^{-2}$, relative to the continuum intensity. The white box indicates the area where inversions are carried out.}
        \label{fig:synthetic_stokes}
\end{figure}

Figure~\ref{fig:synthetic_stokes} shows the resulting Stokes~$I$, $Q$, $U$ and $V$ images at $-209$ m\AA \ from the core of the \ion{Ca}{ii}~$\lambda8542$ \AA \ line.

The synthetic spectral profiles are convolved with Gaussian instrumental profiles of full-width-half-maximum $\mathrm{FWHM}=50$~m\AA. Additive photon noise is added afterwards with standard deviations $\sigma=[10^{-4}, 10^{-3}]$.

\begin{figure}[]
      \centering
        \resizebox{\textwidth}{!}{\includegraphics[]{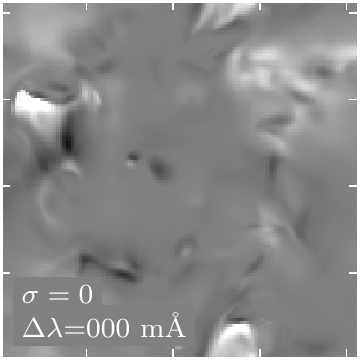}\includegraphics[]{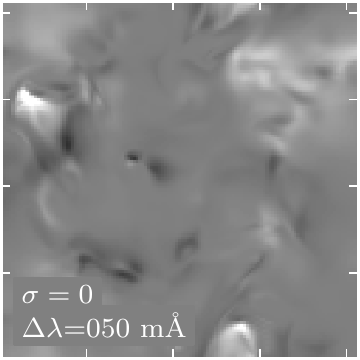}\includegraphics[]{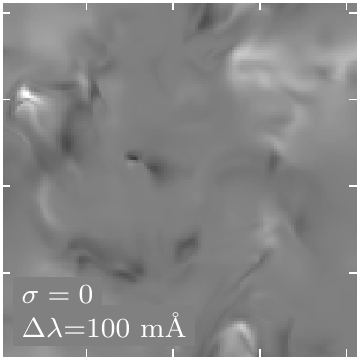}}
        \resizebox{\textwidth}{!}{\includegraphics[]{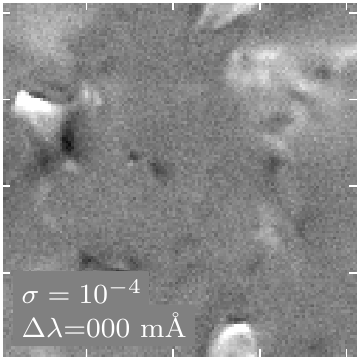}\includegraphics[]{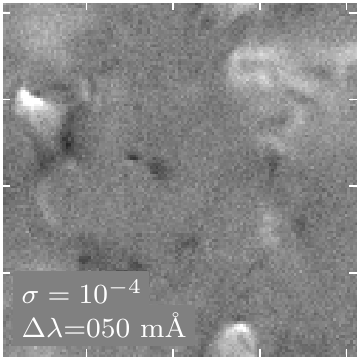}\includegraphics[]{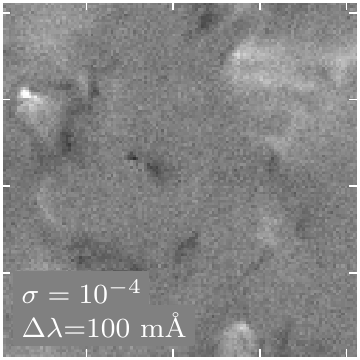}}
         \resizebox{\textwidth}{!}{\includegraphics[]{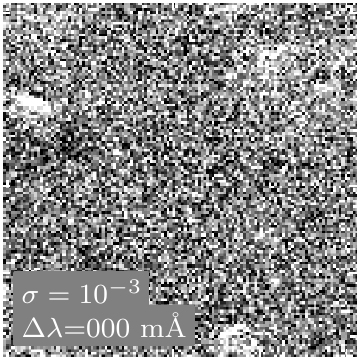}\includegraphics[]{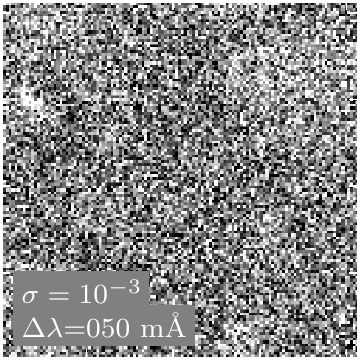}\includegraphics[]{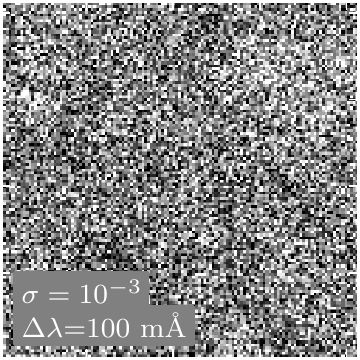}}
        \caption{Synthetic Stokes Q images computed at $+205$~m\AA \  from the core of the \ion{Ca}{ii}~$\lambda8542$~\AA \ line. The spectra have been degraded with Gaussian profiles and random photon noise. From left to right, the full-width-half-maximum of the Gaussian profile increases from $\mathrm{\Delta\lambda}=0$~m\AA \ (no convolution) to $\mathrm{\Delta\lambda}=100$~m\AA. From top to bottom, the noise is increased from $\mathrm{\sigma}=0$ to $\mathrm{\sigma}=10^{-3}$. The images are scaled to $\pm0.15\%$, relative to the continuum intensity.}
        \label{fig:stokesQ}
\end{figure}

In Fig.~\ref{fig:stokesQ}, the Stokes~$Q$ image from the central part of our snapshot is shown at $+205$~m\AA \  from the line core. The profiles have been convolved with Gaussian instrumental profiles of full-width-half-maximum (FWHM hereafter) $\mathrm{\Delta\lambda} = 50$~m\AA. Random photon noise has been added with standard deviation $\mathrm{\sigma} = [0,10^{-4},10^{-3}]$. Most spectral features in Stokes~$Q$ and $U$ have signals of the order of $10^{-4}$, peaking at most at $10^{-3}$ where the magnetic field is strong.

In the middle row, corresponding to $\sigma=10^{-4}$, most polarization features are above the noise level, regardless of the spectral resolution. However, in the bottom row ($\sigma=10^{-3}$), the limited spectral resolution and noise wash all the features from the field of view.

From an observational perspective, the magnetic field orientation and inclination are computed using Stokes~$Q$ and $U$. To infer the full magnetic field vector, it is crucial to carry out observations with noise well below $10^{-3}$, as it is shown in Sect.~\ref{sec:inversions}.

\section{Inversions}\label{sec:inversions}

The inversions are carried out using the inversion code \textsc{Nicole}. The results shown in this section correspond to chromospheric averages computed from $\mathrm{Log}(\tau_{500})=-4$ to $\mathrm{Log}(\tau_{500})=-6$.

The inversions are initialized using the VAL-C model \citep{1981ApJS...45..635V}. During the inversion, the variables of the model are modified at node points placed equidistantly along the depth scale, to fit the observed profiles.  

Each inversion is computed in two cycles, a similar approach to that described by \citet{2007ApJ...670..885P,2011A&A...529A..37S}. The first cycle aims to search for a rough solution using few node points. In the second cycle more degrees of freedom are allowed to refine the solution. The inversions are computed using critically sampled spectral profiles, assuming that the spectral resolution is given by the FWHM of the instrumental profile. 

\begin{figure}[]
      \centering
        \resizebox{\textwidth}{!}{\includegraphics[]{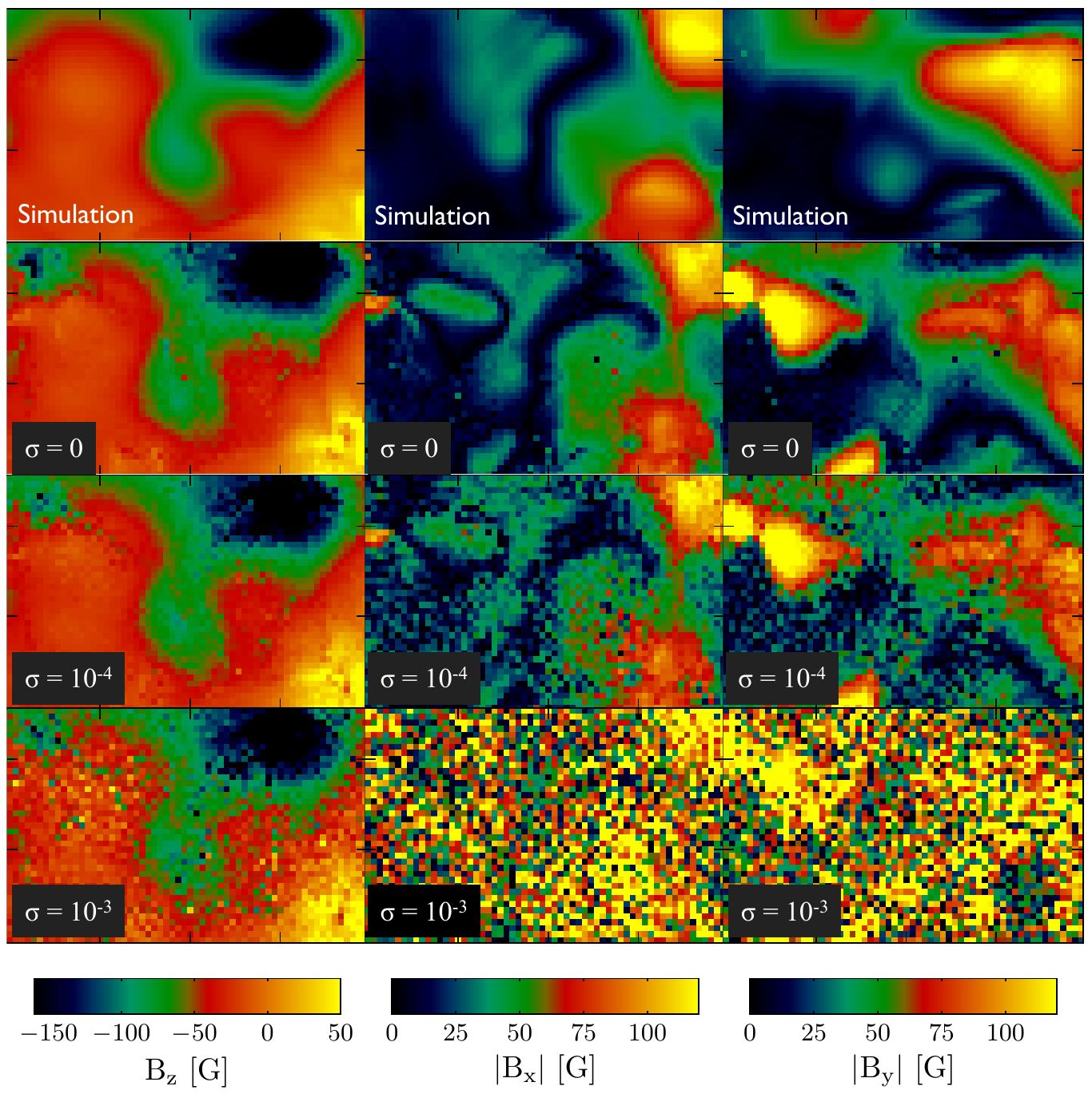}}
        \caption{Comparison of the magnetic field vector from the 3D snapshot and the results from the inversion. Gaussian instrumental profiles with $\mathrm{FWHM}= 50$~m\AA \ have been used. From left to right, the panels show $\mathrm{B}_z(x,y), |\mathrm{B_x}(x,y)|, |\mathrm{B_y}(x,y)|$. From top to bottom, the magnetic field from the 3D snapshot, followed by the inversion results assuming random photon noise of $\sigma=0, 10^{-4}, 10^{-3}$ correspondingly.}
        \label{fig:inv_panels}
\end{figure}

The results from the inversions are shown in Fig.~\ref{fig:inv_panels}. The reconstructed $\mathrm{B}_z$ is less noisy than $|\mathrm{B}_x|$ and  $|\mathrm{B}_y|$, an expected results given that $\mathrm{B}_z$ is determined from Stokes~$V$, where signals are much stronger than in Stokes~$Q$ and $U$. 

Noise is a critical factor for the inversions. The inversion does a reasonably good job reconstructing $\mathrm{B}_z$ for all the noise levels considered in this work. The differences appear with $|\mathrm{B}_x|$ and  $|\mathrm{B}_y|$. The results obtained with $\sigma=10^{-4}$ are similar to the noise-free case where there is good correspondence between the inversion and the 3D simulation. At $10^{-3}$, all the magnetic features are lost, under the noise level. 


\begin{figure}[]
      \centering
        \resizebox{0.6\textwidth}{!}{\includegraphics[]{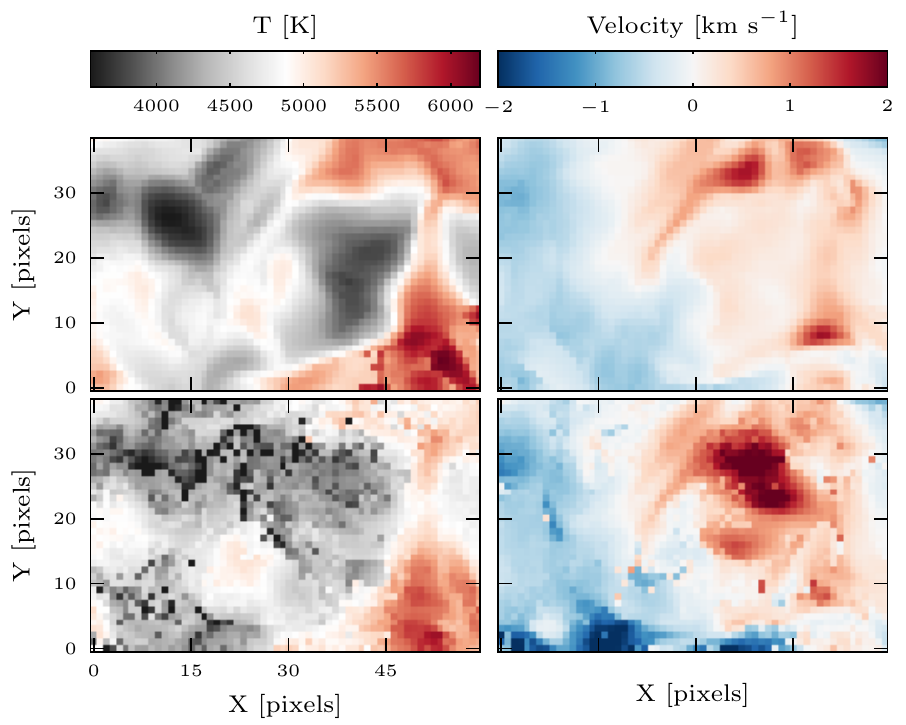}}
        \caption{The results from the NLTE inversion are compared with the quantities present in the 3D simulation. The top row corresponds to temperature and line-of-sight velocity from the 3D model, and the bottom row to the inversion results. These panels correspond to a chromospheric average from $\mathrm{Log}(\tau_{500})=-4$ to $\mathrm{Log}(\tau_{500})=-6$.}
        \label{fig:tempvel}
\end{figure}

The inverted temperature and line-of-sight velocity (l.o.s. hereafter) are insensitive to the spectral degradation and noise levels considered here, thus only one panel is shown in Fig.~\ref{fig:tempvel}. The l.o.s. velocity is recovered mostly from the Stokes~$I$ and $V$ profiles. The inversion shows similar feature to those in the 3D model, with minor differences probably due to differences in the quality of the fit and the rough interval that has been selected to compute the depth averages.

The inferred temperature map has less contrast than the 3D simulation. We expect 3D NLTE effects to work in the same direction. \citet{2009ApJ...694L.128L} reports decreased contrast in intensity monochromatic images computed in 3D NLTE when compared with the 1.5D case. This effect would lead to decreased contrast in temperature when inverting those profiles, however it is hard to quantify by how much. The hydrostatic equilibrium assumption does not seem to have a large impact on the inverted quantities.

\section{Conclusions}
In this paper we present preliminary results from an ongoing work that will be presented in detail in a forthcoming publication. 

We have created synthetic \ion{Ca}{ii}~$\lambda8542$ observations in NLTE, evaluating the 3D radiation field and assuming Zeeman induced polarization. The resulting profiles have been degraded with a list of instrumental profiles and noise and inverted afterwards. To infer the full magnetic field vector from observations, the noise level must be of the order of $10^{-4}$ and clearly below $10^{-3}$. 

The inversions are computed assuming plane-parallel geometry and imposing hydrostatic equilibrium. Despite these assumptions, the NLTE inversion is able to successfully reconstruct the chromospheric quantities from the 3D simulation, although the quality of $|\mathrm{B}_x|$ and  $|\mathrm{B}_y|$ strongly depends on noise. 

The 3D NLTE effects mentioned in Sec.~\ref{sec:inversions}, tend to decrease the contrast of intensity profiles. The decreased contrast in the inverted temperature map could be explained with this effect.

\acknowledgements The authors of this work thank Luc Rouppe van der Voort for helping with the data reduction of the observations shown in this paper. JdlCR acknowledges financial support from Nikolai Piskunov to attend this meeting. 

\bibliographystyle{asp2010}
\bibliography{delacruz}

\end{document}